\documentclass[a4paper, 11pt, oneside]{article} 
\usepackage{geometry}
\usepackage{booktabs} 
\usepackage{mathptmx} 
\usepackage{fancyhdr}
\usepackage[normalem]{ulem}
\usepackage[utf8]{inputenc}
\usepackage{microtype}
\usepackage{graphicx}
\usepackage{colortbl}
\usepackage{multirow}
\usepackage{amsmath}
\usepackage{bm}
\usepackage{epsfig}
\usepackage{authblk}
\usepackage{amssymb,enumitem}
\usepackage[hyphens]{url}
\usepackage{hyperref}
\usepackage[hypcap = false]{caption}
\usepackage{color}
\usepackage{soul}
\usepackage{bbding}
\usepackage{graphicx, subfig}
\usepackage{listings}
\usepackage{fontawesome}
\usepackage{longtable}
\usepackage{algorithm}
\usepackage{algorithmic}
\usepackage{mdsymbol}
\usepackage{color}

\newcommand{\tabincell}[2]{\begin{tabular}{@{}#1@{}}#2\end{tabular}}

\def\lara{\textsc{Lara}}
\def\iterlara{\textsc{IterLara}}
\definecolor{mygray}{gray}{.88}

\begin{document} 

\begin{titlepage} 

	\centering 
	
	\scshape 
	
	\vspace*{\baselineskip} 
	
	
	\rule{\textwidth}{1.6pt}\vspace*{-\baselineskip}\vspace*{2pt} 
	\rule{\textwidth}{0.4pt} 
	
	\vspace{0.75\baselineskip} 
	
	{\LARGE IterLara: A Turing Complete Algebra for Big\\Data, AI, Scientific Computing, and Database} 
	
	\vspace{0.75\baselineskip} 
	
	\rule{\textwidth}{0.4pt}\vspace*{-\baselineskip}\vspace{3.2pt} 
	\rule{\textwidth}{1.6pt} 
	
	\vspace{2\baselineskip} 
	
	
	
	\vspace*{3\baselineskip} 
	
	
	Edited By
	
	\vspace{0.5\baselineskip} 
	
	{\scshape\Large Hongxiao Li\\ Wanling Gao\\ Lei Wang\\Jianfeng Zhan\\ }
	
	
	\vspace{0.5\baselineskip} 

	\vfill 
	
	
	\epsfig{file=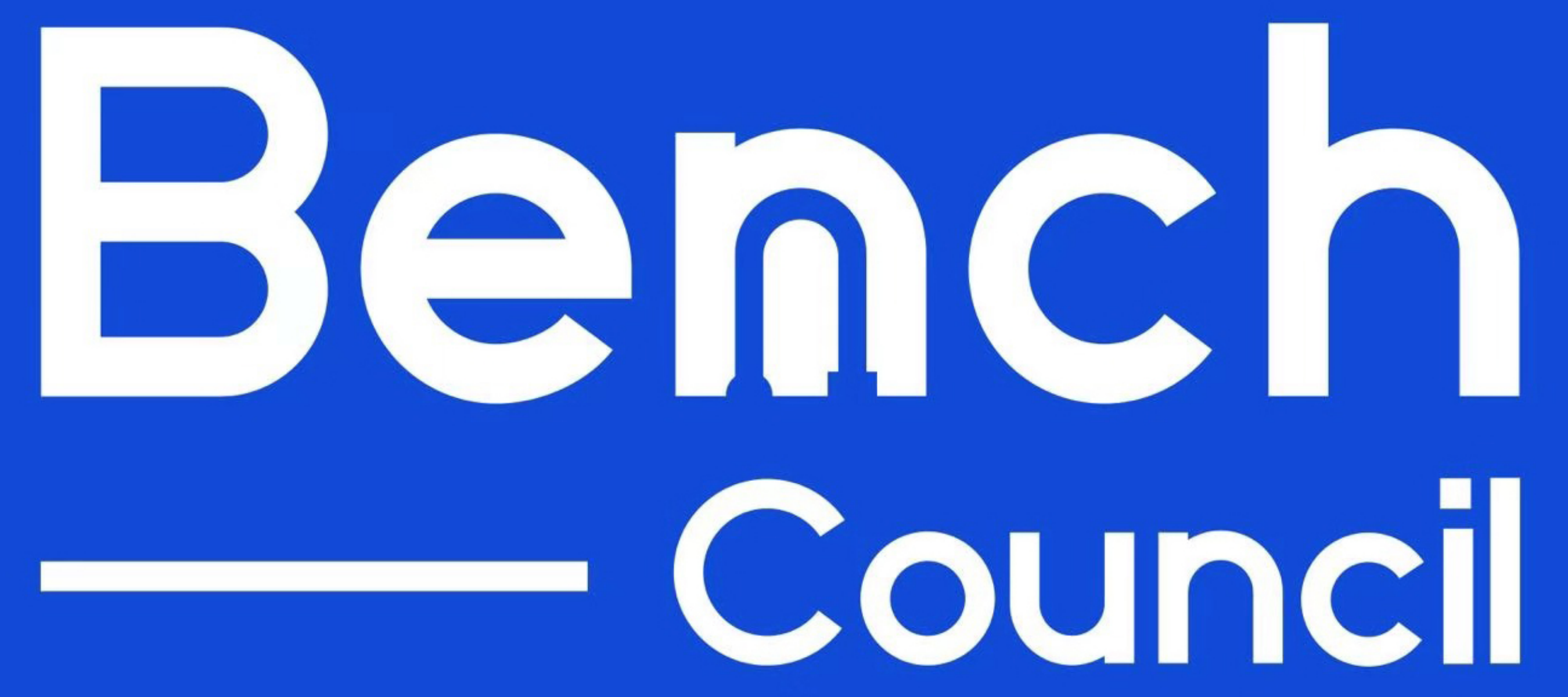,height=2cm}
	\textit{\\BenchCouncil: International Open Benchmark Council\\Chinese Academy of Sciences\\Beijing, China\\https://www.benchcouncil.org} 
	\vspace{5\baselineskip} 

	Technical Report No. BenchCouncil-IterLara-2023 
	
	{\large Jun 20, 2023} 

\end{titlepage}


\title{IterLara: A Turing Complete Algebra for Big Data, AI, Scientific Computing, and Database}

\author[1,3]{Hongxiao Li}
\author[1,2,3]{Wanling Gao}
\author[1,2,3]{Lei Wang}
\author[1,2,3]{Jianfeng Zhan\thanks{Jianfeng Zhan is the corresponding author.}}

\affil[1]{Research Center for Advanced Computer Systems, State Key Lab of Processors, Institute of Computing Technology, Chinese Academy of Sciences\\ \{gaowanling, wanglei\_2011, zhanjianfeng\}@ict.ac.cn}
\affil[2]{BenchCouncil (International Open Benchmark Council)}
\affil[3]{University of Chinese Academy of Sciences\\ lihongxiao19@mails.ucas.ac.cn}

\date{Jun 20, 2023}
\maketitle

\begin{abstract}

\textsc{Lara} is a key-value algebra that aims at unifying linear and relational algebra with three types of operation abstraction. The study of \textsc{Lara}'s expressive ability reports that it can represent relational algebra and most linear algebra operations. However, several essential computations, such as matrix inversion and determinant, cannot be expressed in \textsc{Lara}. \textsc{Lara} cannot represent global and iterative computation, either.  This article proposes \textsc{IterLara}, extending \textsc{Lara} with iterative operators, to provide an algebraic model that unifies operations in general-purpose computing, like big data, AI, scientific computing, and database. We study the expressive ability of \textsc{Lara} and \textsc{IterLara} and prove that \textsc{IterLara} with aggregation functions can represent matrix inversion, determinant. Besides, we demonstrate that \textsc{IterLara} with no limitation of function utility is Turing complete. We also propose the Operation Count (OP) as a metric of computation amount for \textsc{IterLara} and ensure that the OP metric is in accordance with the existing computation metrics.

\end{abstract}

\section{Introduction}

Good data and computation abstractions catch the intrinsic properties of computing and hence play an essential role, such as linear algebra in the scientific computing field and relational algebra in the database field. Researchers try to propose a concise and unified algebraic model with the trend of computation fusion of big data, AI, scientific computing, and database fields. Unfortunately, this is not a trivial issue.

Researchers propose different data and computation abstractions in other fields, such as tensor computation graphs in the AI field and relation graphs in the database fields. Unfortunately, these abstractions have a different algebraic structure with no equivalent parts with each other. Unifying them, regardless of their algebraic properties, is not feasible. For example, there is no equivalent tensor abstraction in the database field. Besides this, the existing general-purpose models like the Turing machine lack high-level abstractions for computation representation and have less advantage in modeling contemporary computations.

Researchers presented a minority of concise and unified algebraic models for different fields in recent years. A representative model is \lara. \lara~\cite{hutchison2016lara} is an algebraic model that aims to unify linear and relational algebra. It provides a unified data abstraction for arrays, graphs, and tensors named associative table. \lara~can represent all operations in the extended relational algebra field and most operations in the linear algebra field. However, the studies on \lara's expressive ability~\cite{barcelo2022expressiveness,barcelo2019expressiveness} reported that it could not represent several essential computations, such as matrix inversion and determinant. It cannot represent other non-local computations, either.

Our work aims to propose a concise and unified algebraic model that covers big data, AI, scientific computing, and database based on \lara. The algebraic model should be concise in different fields and Turing complete. Our insight is to add minimum operators to \lara~and extend its expressive ability. Since iterative computation is necessary for general-purpose models, we add an iterative operator to \lara~and proposed \iterlara. This trial is also inspired by the Geerts et al.~\cite{geerts2021expressive} as an extension to Matlang~\cite{brijder2019matlang}. We argue that \iterlara~is suitable as a concise and Turing complete algebraic model that unifies computations in big data, AI, scientific computing, and database fields. Table~\ref{tab_contribution} is the brief abstract of \iterlara's expressive ability in comparison to those of the extended relation algebra (RA), for-Matlang~\cite{geerts2021expressive}, and \lara. The result reports that \iterlara~can represent the most computations from different fields while the others cannot. This table includes major operators in big data, AI, scientific computing, and database fields.

Our main contribution is divided into the following four parts:

First, we thoroughly study the expressive ability of the original \lara~(also noted as $\lara(\Omega_{All[n]}^{All})$, it will be explained in late sections), rather than tame-\lara~or other extensions. We proved that matrix inversion ($Inv$) and determinant ($Det$) of an arbitrary matrix could not be represented in \lara. In contrast, \lara~can represent those of a matrix of a specific size ($Inv[n]$ and $Det[n]$). We also present pooling~\cite{gholamalinezhad2020pooling} (e.g., MaxPool, AvgPool) and some other representation in \lara.

Second, we propose a non-trivial extension: iteration, to \lara, and we name a new algebraic model, \iterlara. We consider any computation an expression and describe an iterative operation ($\mathbf{iter}$) that inputs and outputs expression. We present the strict syntax and semantics of all expressions in \iterlara.

Third, we study the expressive ability of \iterlara~in three aspects. We proved that \lara~with constant iteration times (noted as $\iterlara^{Const}$) is equivalent to \lara. We demonstrated that \lara~with aggregation iteration times (indicated as $\iterlara^{Count}$) can represent matrix inversion and determinant~\cite{greub2012linear}. We proved that \iterlara~with no limitation (noted as \iterlara) is Turing complete by constructing representations of all operators of another Turing complete algebra: BF language~\cite{mathis2011BF}. BF language is a minimum Turing complete programming language comprising only eight operators. It is not designed for practical use but for research on language's extreme expressive ability.

Last, we define a computable theoretical metric of computation: Operation Count (OP, in short) that is compatible with the existing computation metrics, such as FLOPs~\cite{institute1985ieee} in the high-performance computing field and time/space complexity in the algorithm analysis field.

The rest of the paper is organized as follows: Section 2 is the related work. Section 3 introduces \lara, including the strict definition of \lara~operations with important examples. Section 4 is the definitive ability study of \lara, including the studies of matrix inversion ($Inv$), determinant ($Det$), and pooling. Section 5 is the definition of \iterlara. Section 6 is the expressive ability study of \iterlara, including the studies of $\iterlara^{Const}$, $\iterlara^{Count}$, and \iterlara. Section 7 is the definition of Operation Count (OP). Section 8 is the conclusion.

\begin{table}[ht]
\centering
\resizebox{0.85\textwidth}{!}{
\begin{tabular}{|cccccc|}
\hline
Computation& Field&\tabincell{c}{The extended\\relational algebra~\cite{aho1979universality}}& for-Matlang~\cite{geerts2021expressive}& \lara~\cite{hutchison2016lara}& \tabincell{c}{\iterlara\\(Ours)}\\\hline
$\sigma,\pi,\rho,\gamma,\cup,\cap$& \tabincell{c}{Big data,\\database}& $\checkmark$& $\times$& $\checkmark$& $\checkmark$\\\hline
$-,\div, \bowtie,\triangleright,\triangleleft$& \tabincell{c}{Big data,\\database}& $\checkmark$& $\times$& $\checkmark$& $\checkmark$\\\hline
$+,\times$& \tabincell{c}{AI, scientific\\computing}& $\checkmark$& $\checkmark$& $\checkmark$& $\checkmark$\\\hline
$gemm,smm$& \tabincell{c}{AI, scientific\\computing}& $\times$& $\checkmark$& $\checkmark$& $\checkmark$\\\hline
\tabincell{c}{pooling\\($MaxPool,\cdots$)}& AI& $\times$& $\checkmark$& $\checkmark$& $\checkmark$\\\hline
\tabincell{c}{activation\\($ReLU,SeLU,\cdots$)}& AI& $\times$& $\times$& $\checkmark$& $\checkmark$\\\hline
\tabincell{c}{$ArgMax$,\\$ArgMin$}& AI, big data& $\checkmark$& $\times$& $\checkmark$& $\checkmark$\\\hline
$Conv$& \tabincell{c}{AI, scientific\\computing}& $\times$& $\checkmark$& $\checkmark$& $\checkmark$\\\hline
$Inv,Det$& \tabincell{c}{Scientific\\computing}& $\times$& $\times$& $\times$& $\checkmark$\\\hline
$EinSum$& \tabincell{c}{AI, scientific\\computing}& $\times$& $\checkmark$& $\checkmark$& $\checkmark$\\\hline
\tabincell{c}{aggregation\\($Count,Sum,\cdots$)}& Database& $\checkmark$& $\times$& $\checkmark$& $\checkmark$\\\hline
First-order logic& \tabincell{c}{AI, big data,\\database}& $\times$& $\times$& $\checkmark$& $\checkmark$\\\hline
Iteration& AI, big data& $\times$& $\times$& $\times$& $\checkmark$\\\hline
\tabincell{c}{for-loop,\\while-loop}& AI, big data& $\times$& $\times$& $\times$& $\checkmark$\\\hline
\end{tabular}}
\caption{This table gives a summary of comparing \iterlara's expressive ability against the extended relation algebra (RA)~\cite{aho1979universality}, for-Matlang~\cite{geerts2021expressive}, and \lara, indicating \iterlara~ has better expressive ability.}
\label{tab_contribution}
\end{table}
\section{Related work}
Tensor computation graphs in the AI field, matrix computation in the scientific computing field, and relational algebra~\cite{aho1979universality} in the database field are the most common algebraic models. There are no unified algebraic models for computation in the big data field.

Our work proposes a unified computation model covering big data, AI, scientific computing, and database. Similar research on unified computation models includes \lara~\cite{hutchison2016lara} and Matlang~\cite{brijder2019matlang}. Table~\ref{tab_contribution} shows the concrete relation between the existing research and ours.

\lara~is a minimalist kernel for a linear and relational algebra database proposed in the LaraDB project~\cite{hutchison2017laradb}. \lara~uses key-value pairs for data abstraction, which can be converted to relation graphs. Barcel{\'o} et al.~\cite{barcelo2022expressiveness,barcelo2019arxiv} concluded that tame-\lara~cannot represent matrix inversion and convolution and proposed an extension to \lara~and it can represent convolution and other local operations. Meanwhile, \lara~is equivalent to first-order logic with aggregation.

The Matlang language is a recently proposed algebraic model as an abstraction. Geerts et al.~\cite{geerts2021expressive} added iterative properties to the Matlang language. Barcel{\'o} et al.~\cite{barcelo2019expressiveness} concluded that Matlang has a lower expressive ability than \lara. This paper inspired us to add iterative properties to other algebraic models. The \iterlara~extension to \lara~is similar to for-Matlang extension to Matlang but has a more extraordinary expressive ability.

\section{The original \lara~language}

This section introduces the definitions, algebraic structures, and semantics of the original \lara~language. We added some other notations for convenience, basically notations from relational algebra. We ensure that every notation is well-defined at its first appearance. The entire list of our notation is in Table~\ref{tab_operator}.

\begin{table}[ht]
\centering
\resizebox{0.85\textwidth}{!}{
\begin{tabular}{|cl|}
\hline
Operator& Description\\\hline
$A,B,C,\cdots$& Associative tables\\\hline
$a,b,c,\cdots$& Records\\\hline
$e,e_1,e_2,\cdots$& Expressions\\\hline
$\vec{k_1},\vec{k_2},\cdots$& Keys\\\hline
$\vec{v_1},\vec{v_2},\cdots$& Values\\\hline
$K_A$& Set of names of key attributes of table $A$\\\hline
$V_A$& Set of names of value attributes of table $A$\\\hline
$E_A$& The empty table that has the same key and value attributes as $A$'s\\\hline
$a\oplus b$& The user-definable addition of two records $a$ and $b$\\\hline
$a\otimes b$& The user-definable multiplication of two records $a$ and $b$\\\hline
$\bigoplus\limits_{x\in A} x,\bigotimes\limits_{x\in A} x$& Aggregation on table $A$\\\hline
$A\hourglass_\oplus B$& The union of table $A$ and $B$, noted as $A\hourglass B$ if addition is not important\\\hline
$A\hat{\bowtie}_\otimes B$& \tabincell{l}{The strict join of table $A$ and $B$, noted as $A\hat{\bowtie}B$ if multiplication is not\\important}\\\hline
$A\bowtie_\otimes B$& \tabincell{l}{The relaxed join of table $A$ and $B$, which is equivalent to natural join ($\bowtie$) in\\relational algebra when $\otimes$ takes the Cartesian product}\\\hline
$\mathbf{ext}_f A$& The extension of function $f$ onto a table $A$\\\hline
$\mathbf{iter}_F^{cond} A$& \tabincell{l}{The iteration operator with condition expression $cond$ and iteration body $F$\\on a table $A$}\\\hline
$\sigma_F A$& Selection on table $A$ with function $F$ in relational algebra\\\hline
$\pi_S A$& Projection on table $A$ with attribute $S$ in relational algebra\\\hline
$\rho_{x/y} A$& Rename on table $A$ with attribute name from $x$ to $y$ in relational algebra\\\hline
$\gamma_S A$& Aggregation on table $A$ with attribute $S$ in relational algebra\\\hline
$\cup,\cap,-,\div$& Set union, intersection, difference, as the same in relational algebra\\\hline
$\triangleright,\triangleleft$& \tabincell{l}{Left and right antijoin in relational algebra, the result only includes the\\common records with attributes from one side. Namely, $A\triangleright B=A-$\\$\pi_{K_A-K_B,V_A-V_B}(A\bowtie B)$, and $A\triangleleft B=B-\pi_{K_B-K_A,V_B-V_A}(A\bowtie B)$}\\\hline
Boole$[expr]$& The Boolean value of the expression $expr$\\\hline
$\mathbf{OP}(expr)$& The operation count of the expression $expr$\\\hline
\end{tabular}}
\caption{List of operators.}
\label{tab_operator}
\end{table}

\subsection{Preliminaries}
In this paper, $\mathbb{N}[a,b]$ stands for integers between $a$ and $b$, where $a$ and $b$ are positive. An associative table (in short as ``table'' if not confusing) is defined as a key-value table of data. The name of keys and values of table $A$ are noted as $K_A$ and $V_A$. A table has no same keys, which means that data (we call ``records'') of the same keys must have the same value. We note the key and the value of record $a$ as $k(a)$ and $v(a)$. Due to the non-replicability of keys, we can safely note the value of a record in table $A$ that has a certain key $\vec{k}$ as $A[\vec{k}]$. Namely for record $a=k(a):v(a)=\vec{k}:\vec{v}$, we have $A[\vec{k}]=\vec{v}$.

Next, we present requirements for user-definable binary functions $\oplus$, $\otimes$ that will appear in the following text. First, these functions must input two values and output one value of the same type. Second, $\oplus$ and $\otimes$ must satisfy commutativity and associativity, namely $a\theta b=b\theta a$ and $(a\theta b)\theta c=a\theta(b\theta c)$ where $\theta$ stands for $\oplus$ or $\otimes$. Third, $\oplus$ and $\otimes$ must have an identity element. For example, the identity element for addition is $0$, and the identity element for multiplication is $1$. For all the values in table $A$, an aggregation function is defined as the associative computation on every value in $A$, noted as $\bigoplus A=\bigoplus\limits_{x\in A} x$ and $\bigotimes A=\bigotimes\limits_{x\in A} x$. Symbols $sum$, $min$, $max$, and $count$ are four aggregation functions that stand for summation, minimum, maximum, and total count, respectively. Besides, $avg$ represents the average function, which can be computed with $count$ and $sum$.

We also introduce the symbols in relational algebra. Symbol $\sigma_F A$ stands for selection on table $A$ with function $F$. Symbol $\pi_S A$ stands for projection on table $A$ with attribute $S$. Symbol $\rho_{x/y} A$ stands for rename on table $A$ with attribute name from $x$ to $y$. Symbol $\gamma_S A$ stands for aggregation on table $A$ with attribute $S$. Symbols $\cup$, $\cap$, $-$, $\div$ are set union, intersection, difference, and division, as the same in relational algebra. Symbol $\bowtie$ stands for table join. As the relaxed table join in \lara~and the natural join in relational algebra are equivalent when $\otimes$ takes the Cartesian product, we adopt the same symbol. In the extended relational algebra, symbols $\triangleright$ and $\triangleleft$ are left and right antijoin.

\subsection{Data abstraction}
The only data abstraction of \lara~is the associative table, firstly proposed in~\cite{kepner2016associative}. We use $A:[\vec{k}:\vec{v}]$ to denote an associative table with $\vec{k}=[k_1,\cdots,k_n]$ attributes and $\vec{v}=[v_1,\cdots,v_m]$. For example, a matrix $M$ can be represented as a table: $M:[[i,j]:v]$, where $[i,j]$ is the key that stands for the coordinate and $v$ is the value that stands for the data. Besides matrices and tensors, a relation table in relational algebra can also be represented as a table, where a key is a unique id and a value is the data in a record. The symbol $E_A$ represents an empty table with the same key and value attributes as $A$'s.

An association table must include a finite quantity of records. The value attribute of each record must have limited support; namely, it must take value from a limited amount of data, and the data must include a default value. The default value is the identity value of aggregation functions, and it is a zero-element or a one-element in most conditions with no specification.

\subsection{Operators}
\lara~includes three types of operators: union, join, and ext, parameterized by user-definable ``sum'' ($\oplus$), ``multiply'' ($\otimes$), and ``flatmap'' ($f$) functions, respectively. The join operator includes the strict join and the relaxed join. We present the definition of them as follows.

\subsubsection{Table union}
The union operator inputs two tables, $A$ and $B$. With given attributes' formats of $A$ and $B$, the union $A\hourglass_\oplus B$ is defined as follows:\\
For tables $A$ and $B$ that
\begin{equation}A:[a_1,\cdots,a_m,c_1,\cdots,c_n:x_1,\cdots,x_q,z_1,\cdots,z_r]\nonumber; B:[c_1,\cdots,c_n,b_1,\cdots,b_p:z_1,\cdots,z_r,y_1,\cdots,y_s]\nonumber\end{equation}
where $c$'s are common part of keys and $a$'s and $b$'s are individual parts of keys, $z$'s are common part of values and $x$'s and $y$'s are individual parts of values, we have
\begin{equation}A\hourglass_\oplus B:[c_1,\cdots,c_n:x_1,\cdots,x_q,z_1,\cdots,z_r,y_1,\cdots,y_s]\nonumber\end{equation}
where
\begin{equation}\begin{aligned}\Big(A\hourglass_\oplus B\Big)[c_1,\cdots,c_n]=\Big[\bigoplus\limits_{a_1,\cdots,a_m}\pi_x A[a_1,\cdots,a_m,c_1,\cdots,c_n],\bigoplus\limits_{a_1,\cdots,a_m}\pi_z A[a_1,\cdots,a_m,c_1,\cdots,c_n]\\\oplus\bigoplus\limits_{b_1,\cdots,b_p}\pi_z B[c_1,\cdots,c_n,b_1,\cdots,b_p],\bigoplus\limits_{b_1,\cdots,b_p}\pi_y B[c_1,\cdots,c_n,b_1,\cdots,b_p]\Big]\end{aligned}\nonumber\end{equation}

Table union on $A$ and $B$ retains the intersection of $K_A$ and $K_B$ and the union of $V_A$ and $V_B$, with element-wise $\oplus$ aggregation on the data.

\subsubsection{Strict table join}
The strict join operator inputs two tables, $A$ and $B$. With given attributes' formats of $A$ and $B$, the union $A\hat{\bowtie}_\otimes B$ is defined as follows:\\
For tables $A$ and $B$ that
\begin{equation}A:[a_1,\cdots,a_m,c_1,\cdots,c_n:x_1,\cdots,x_q,z_1,\cdots,z_r]\nonumber; B:[c_1,\cdots,c_n,b_1,\cdots,b_p:z_1,\cdots,z_r,y_1,\cdots,y_s]\nonumber\end{equation}
and $K_A\cap V_B=K_B\cap V_A=\emptyset$, we have
\begin{equation}A\hat{\bowtie}_\otimes B:[a_1,\cdots,a_m,c_1,\cdots,c_n,b_1,\cdots,b_p:z_1,\cdots,z_r]\nonumber\end{equation}
where
\begin{equation}\begin{aligned}\Big(A\hat{\bowtie}_\otimes B\Big)[a_1,\cdots,a_m,c_1,\cdots,c_n,b_1,\cdots,b_p]=\pi_z A[a_1,\cdots,a_m,c_1,\cdots,c_n]\otimes\pi_z B[c_1,\cdots,c_n,b_1,\cdots,b_p]\end{aligned}\nonumber\end{equation}

Strict table join on $A$ and $B$ retains the union of $K_A$ and $K_B$ and the intersection of $V_A$ and $V_B$, with element-wise $\otimes$ aggregation on the data.

\subsubsection{Ext}
For tables $A$ that
\begin{equation}A:[a_1,\cdots,a_m:x_1,\cdots,x_n:0_1,\cdots,0_n]\nonumber\end{equation}
where $0_1,\cdots,0_n$ are default values, we first present the requirements for the flatmap function $f$ as follows:
\begin{equation}f:a_1\times\cdots\times a_m\times x_1\times\cdots\times x_n\rightarrow(b_1\times\cdots\times b_{m'}\rightarrow y_1\times\cdots\times y_{n'})\nonumber\end{equation}

This formula means that function $f$ inputs a table of the same format as $A$ and outputs a table of the format $[b_1,\cdots,b_{m'}:y_1,\cdots,y_{n'}:0_1',\cdots,0_{n'}']$, where $0_1',\cdots,0_{n'}'$ are default values that $f([\vec{k_i},0_1,\cdots,0_n])=0_1',\cdots,0_{n'}'$ for any $\vec{k_i}$. Besides this, function $f$ must satisfy another requirement that it has finite support, namely only for a finite quantity of inputs $I$, $f(I)$ is not a default value.

With given attributes' formats of $A$ as the above, the ext operator $\mathbf{ext}_f A$ is defined as follows:
\begin{equation}\mathbf{ext}_f A:[a_1,\cdots,a_m,b_1,\cdots,b_{m'},\rightarrow y_1\times\cdots\times y_{n'}:0_1',\cdots,0_{n'}']\nonumber\end{equation}
where
\begin{equation}\Big(\mathbf{ext}_f A\Big)[a_1,\cdots,a_m,b_1,\cdots,b_{m'}]=f(a_1,\cdots,a_m:A[a_1,\cdots,a_m])[b_1,\cdots,b_{m'}]\nonumber\end{equation}

The ext operator can be considered mapping from a table format to another table format within a given pattern. Especially, when the flatmap function $f$ does not change the table format, namely $m=m'$ and $[a_1,\cdots,a_m]=[b_1,\cdots,b_{m'}]$, we call this operator as map, noted as $map_f A$.

\subsubsection{Relaxed table join}
As we have presented, strict table join on $A$ and $B$ retains the intersection of $V_A$ and $V_B$. Relaxed table join retains the union of $V_A$ and $V_B$ and does not drop any value attributes. Relaxed table join is equivalent to the natural join in relational algebra when $\otimes$ takes the Cartesian product. In detail, we have $A\bowtie_\otimes B=\mathbf{ext}_{f_A}\mathbf{ext}_{f_B}(A\hat{\bowtie}_\otimes B)$, where $f_A$ is a identity mapping that changes the value attribute $V_A$ to $V_A\cup V_B$, and $f_B$ is similar. The data at the extended attributes take the default value.

\subsection{Examples}
To make it clearer, we present some examples of \lara~computation.\\
\textbf{Example}:\\
We define associative tables $A$ and $B$ in Tables~\ref{tab_A} and~\ref{tab_B}.\\

\begin{minipage}[c]{0.5\textwidth}
\centering
\begin{tabular}{cc|cc}
a& c& x& z\\\hline
0& 0& 1& 2\\
0& 1& 2& 4\\
1& 0& 3& 6\\
1& 1& 4& 8\\
\end{tabular}
\captionof{table}{Associative table $A$}
\label{tab_A}
\end{minipage}
\begin{minipage}[c]{0.5\textwidth}
\centering
\begin{tabular}{cc|cc}
c& b& z& y\\\hline
0& 0& 1& 7\\
0& 1& 3& 5\\
1& 0& 5& 3\\
1& 1& 7& 1\\
\end{tabular}
\captionof{table}{Associative table $B$}
\label{tab_B}
\end{minipage}

As examples, the results of $A\hourglass_{max}B$, $A\bowtie_\times B$ (relaxed join), and $\mathbf{ext}_{3_w} A$ is shown in Tables~\ref{tab_AunionB}, \ref{tab_AjoinB}, and~\ref{tab_extA}. Note that $3_w$ stands for a constant function with value three on an attribute $w$.\\
\begin{minipage}[c]{0.33\textwidth}
\centering
\begin{tabular}{c|ccc}
c& x& z& y\\\hline
0& 3& 6& 7\\
1& 4& 8& 5\\
\end{tabular}
\captionof{table}{$A\hourglass_{max}B$}
\label{tab_AunionB}
\end{minipage}
\begin{minipage}[c]{0.33\textwidth}
\centering
\begin{tabular}{ccc|ccc}
a& c& b& x& z& y\\\hline
0& 0& 0& 1& 2& 7\\
0& 0& 1& 1& 6& 5\\
0& 1& 0& 2&20& 3\\
0& 1& 1& 2&28& 1\\
1& 0& 0& 3& 6& 7\\
1& 0& 1& 3&18& 5\\
1& 1& 0& 4&40& 3\\
1& 1& 1& 4&56& 1\\
\end{tabular}
\captionof{table}{$A\bowtie_\times B$}
\label{tab_AjoinB}
\end{minipage}
\begin{minipage}[c]{0.33\textwidth}
\centering
\begin{tabular}{cc|c}
a& c& w\\\hline
0& 0& 3\\
0& 1& 3\\
1& 0& 3\\
1& 1& 3\\
\end{tabular}
\captionof{table}{$\mathbf{ext}_{3_w} A$}
\label{tab_extA}
\end{minipage}

\section{\lara's expressive ability}
In this section, we thoroughly study the expressive ability of \lara~and compare it with some extended \lara~algebra. Before the study, we illustrate some symbols of algebra. The symbol $\lara(S_s^t)$ stands for the \lara~language, with the following restrictions: first, all the data is within the range of $S$; second, only operators in the set $s$ can be used on the key attributes; third, only operators in the set $t$ can be used on the value attributes. The last two restrictions limit choices on the user-definable functions.

Hutchison et al.~\cite{hutchison2016lara} introduced the \lara~language without restrictions on the user-definable functions. However, according to the original study in~\cite{hutchison2016lara}, the semantics should be that the user-definable functions must have a certain finite quantity $n$ of input parameters. Therefore, we note the original \lara~as $\lara(\Omega_{All[n]}^{All})$, where $\Omega$ stands for the total set of data, without any restriction on value range. Hutchison et al.~\cite{hutchison2016lara} proved that the relational algebra ($RA$) can be represented in \lara, namely $RA\in\lara(\Omega_{All[n]}^{All})$. Hutchison et al.~\cite{hutchison2016lara} also proved that general matrix multiplication ($gemm$) and the sparse matrix multiplication ($smm$) can be represented in \lara, namely $gemm,smm\in\lara(\Omega_{All[n]}^{All})$.

Barcel{\'o} et al.~\cite{barcelo2022expressiveness,barcelo2019arxiv} proposed a restricted version of \lara, named tame-\lara, which we note as $\lara(\Omega_=^{All})$. We have $\lara(\Omega_=^{All})\in\lara=\lara(\Omega_{All[n]}^{All})$. They proved that the extended relational algebra ($RA_{Agg}$) can be represented in tame-\lara, where $Agg$ means the aggregation functions added. Barcel{\'o} et al.~\cite{barcelo2022expressiveness,barcelo2019arxiv} also proposed that \lara~can represent select ($\sigma$), project ($\pi$), rename ($\rho$), union ($\cup$), intersection ($\cap$), difference (-), division ($\div$), aggregation ($\gamma$), and other relational operations. We will use relational algebraic symbols instead of \lara~expressions for simplicity in our late text on expressive ability study. They also proved that $Inv,Conv\notin\lara(\Omega_=^{All})$, where $Inv$ and $Conv$ stand for matrix inversion and convolution, and \lara~cannot represent iteration. They proposed an extension of \lara, we note as $\lara(\Omega_<^{\{+,\times\}})$. They proved that $Conv\in\lara(\Omega_<^{\{+,\times\}})$. Barcel{\'o} et al.~\cite{barcelo2019expressiveness} proved that $Einsum,Conv\in\lara(\Omega_{All[n]}^{All})$, where $Einsum$ stands for the Einstein summation~\cite{einsum}.

Our contribution to \lara's expressive ability study is that we prove $Inv\notin\lara(\Omega_{All[n]}^{All})$ and $Inv[n],Pool\in\lara(\Omega_{All[n]}^{All})$, where $Pool$ stands for pooling. We also represent some other important computations. A brief illustration of conclusions is shown in Table~\ref{tab_conclusion}.\\
\subsection{\lara~cannot represent matrix inverse}
We prove that $Inv\notin\lara(\Omega_{All[n]}^{All})$ in two steps.\\
\textbf{Proposition}:\\
A matrix inverse is representable in \lara~only if the determinant is representable in \lara.\\
\textbf{Proof}:\\
Consider the adjugate matrix $A^*$ is representable in \lara~as the following:\\
The algebraic minor matrix is\\
$m_{x,y}B=\begin{cases}(-1)^{i+j}b_{i,j},\mbox{ if }i\neq x\mbox{ and }j\neq y\\0\mbox{ (default), otherwise}\end{cases}$.\\
Therefore, $A^*=\hourglass_\cup^{i,j}A_{i,j}=\hourglass_\cup^{i,j}\mathbf{ext}_{m_{i,j}}A$.\\
Define matrix division function $\dfrac AB=\dfrac{a_{1,1}}{b_{1,1}}$ with the requirement of $A$ and $B$ as the following:\\
$\forall i,j,x,y,\dfrac{a_{i,j}}{b_{i,j}}=\dfrac{a_{x,y}}{b_{x,y}}$.\\
Matrix division is defined only when $A$ is constant times of $B$.\\
According to linear algebra, the adjugate matrix $A^*=Det(A) Inv(A)$ if A is invertible.\\
If $Inv(A)$ can be represented as $I(A)$, since $A$ is invertible, then $Det(A)=\dfrac{A^*}{I(A)}$ is representable in \lara. So matrix inverse is representable in \lara~only if the determinant is representable in \lara.\hfill{$\square$}\\
\textbf{Proposition}:\\
Determinant is not representable in \lara.\\
\textbf{Proof}:\\
Suppose $Det(A)$ can be represented as $f(A)$, where $A\in\mathbb{R}^{n\times n}$. For convenience, we suppose $n>3$.\\
Consider $f'(A)\overset{\mbox{def}}{=}\mathbf{ext}_f A_{[a_1,\cdots,a_n]}$, where $A_{[a_1,\cdots,a_n]}$ is a matrix that has 1 value at $A_{1,a_1},\cdots,$\\$A_{n,a_n}$ and 0 value at other entries, where $a_1,\cdots,a_n\in\mathbb{N}[1,n]$. Because $\{A_{a_1,\cdots,a_n}\}\subseteq\mathbb{R}^{n\times n}$, $f'(A)$ is representable.\\
Consider $g(A)=\mbox{Boole}[\mathbf{ext}_{f'} A]$, where Boole$[expr]$ equals the Boolean value of $expr$. According to linear algebra, $g(A)$  equals $1$ value when $a_1,\cdots,a_n$ are pairwise unequal and equals 0 otherwise. In other words,
$g(A)=\begin{cases}1,\mbox{ if }\bigwedge_{i,j\in[1,n],i\neq j}\mbox{Boole}[a_i\neq a_j]\\0,\mbox{ otherwise}\end{cases}$.\\
Because $B(x,y)\overset{\mbox{def}}{=}\mbox{Boole}[x\neq y]$ does not satisfy transitivity, namely $B(y,z)$ is independent of $B(x,y)$ and $B(x,z)$ for arbitrary $x,y,z$, $g(A)$ must be a function of the total set $a_i$. We note that $g(A)\overset{\mbox{def}}{=}g'(<a_1,\cdots,a_n>)$.\\
Function $g'$ has a total of n independent variables. Consider the three \lara~operations. Table union only accepts union with a unary key attribute, rather than with $<a_1,\cdots,a_n>$. Table join does not do aggregation on value attributes. Therefore, $g'$ representation must include flatmap operation. Construct function $g_0':<a_1,\cdots,a_n>\times\#_1\rightarrow(\#_2\rightarrow\#_3)$, where $\#_i,i=1,2,3$ stands for dummy variables.\\
Since $n$ can be an integer bigger than any finite range, $g_0'$ must have the transitive closure of an infinite number of independent variables, which is impossible.\hfill{$\square$}

According to the above proof, matrix inverse is not representable in \lara.\\
\subsection{matrix inversion at a specific size n}
While a general matrix inversion cannot be represented, we present the representation of inversion to a particular size $n$.\\
\textbf{Proposition}:\\
$Inv[n]\in\lara(\Omega_{All[n]}^{All})$.\\
\textbf{Proof}:\\
Given a matrix sized $n$, represented as table $A:[[i,j]:v]$. According to linear algebra, the determinant equals the sum of the product of a matrix's inverse order number and the values of the corresponding entries (coordinates). Let $n!$ functions $h_1,\cdots,h_{n!}$ be the full permutations of indexes $1,\cdots,n$, where $h_1=[1,2,\cdots,n]$, $h_{n!}=[n,n-1,\cdots,1]$, and so on. Let function $\tau$ be the inverse order number in linear algebra. We have $\forall i\in\mathbb{N}[1,n!],\tau(h_i)=\pm1$. Since $n$ is finite, construct tables $B_k:[[j,k]:w],B_k[j,k]=1,\forall k=1,\cdots,n$. We have $(A\bowtie_\times B_1\bowtie_\times B_2\bowtie_\times\cdots\bowtie_\times B_n)[k_1,\cdots,k_n]=A[1,k_1]A[2,k_2]\cdots A[n,k_n]$. Construct function $d$ such that\\
$d(k_1,\cdots,k_n,w)=\begin{cases}\tau(k_1,\cdots,k_n)\times w\mbox{, if }\bigwedge_{i,j\in[1,n],i\neq j}\mbox{Boole}[a_i\neq a_j]\\0,\mbox{ otherwise}\end{cases}$.\\
Construct table $C:[[k_1,\cdots,k_n]:w]$ with only full permutation items. We have that $Det(A)=C\hourglass_+\mathbf{ext}_d(A\bowtie_\times B_1\bowtie_\times B_2\bowtie_\times\cdots\bowtie_\times B_n)$.\\
As is proved that $A^*$ is representable, at a certain size $n$, we proved that $Inv(A)=\dfrac{A^*}{Det(A)}=A^*\bowtie_\times\Big(\mathbf{ext}_{reciprocal}Det(A)\Big)$ can be represented in $\lara(\Omega_{All[n]}^{All})$.\hfill{$\square$}

\subsection{Pooling and other computations}
\textbf{Proposition}:\\
$Pool\in\lara(\Omega_{All[n]}^{All})$.\\
\textbf{Proof}:\\
Consider 1-dimensional average pooling function $AvgPool1d(A,s)$, where $s$ is the stride parameter, and table $A:[i:v]$ is a vector sized $n$. Construct reshaping function $r_s:i\times v\rightarrow(i'\times j\rightarrow v)$ such that $(\mathbf{ext}_{r_s} A)[i',j]=A[i'\times s+j]$, where $i'\in\mathbb{N}[0,i/s-1]$ and $j\in\mathbb{N}[0,s-1]$. The function $r$ reshapes the $i$-sized vector $A$ into a $[i/s,s]$-shaped matrix. Therefore, $AvgPool1d(A,s)=\mathbf{ext}_{avg_j}(\mathbf{ext}_{r_s} A)$, where the average function $avg_j$ aggregates the second index $j$ of the key attribute.\hfill{$\square$}

Table~\ref{tab_A1}, \ref{tab_rA}, and~\ref{tab_avgrA} show an example of pooling computation.

\begin{minipage}[c]{0.33\textwidth}
\centering
\begin{tabular}{c|c}
i& v\\\hline
0& 1\\
1& 3\\
2& 4\\
3& 5\\
4& 7\\
5& 9\\
\end{tabular}
\captionof{table}{$A$}
\label{tab_A1}
\end{minipage}
\begin{minipage}[c]{0.33\textwidth}
\centering
\begin{tabular}{cc|c}
i'& j& v\\\hline
 0& 0& 1\\
 0& 1& 3\\
 1& 0& 4\\
 1& 1& 5\\
 2& 0& 7\\
 2& 1& 9\\
\end{tabular}
\captionof{table}{$\mathbf{ext}_{r_s} A,s=2$}
\label{tab_rA}
\end{minipage}
\begin{minipage}[c]{0.33\textwidth}
\centering
\begin{tabular}{c|c}
i'& v\\\hline
 0& 2  \\
 1& 4.5\\
 2& 8  \\
\end{tabular}
\captionof{table}{$\mathbf{ext}_{avg_j}(\mathbf{ext}_{r_s} A),s=2$}
\label{tab_avgrA}
\end{minipage}

Besides pooling, we present \lara~representation of $ArgMax$ and activation function ($ReLU$) computations on a vector $A$ as follows:\\
$ArgMax(A)=A\div\gamma_{max}A$ (with relational algebraic symbols, for simplicity),\\
$ReLU(A)=\mathbf{ext}_{max_0}A$, where $max_0(x)=max(x,0)$.

Our study mainly focuses on the original \lara. Since pooling, $ArgMax$, and $ReLU$ do not require complicated user-definable functions, they are representable in tame-\lara.

So far, our study reports that \lara~can represent the majority of computation in the AI field. In contrast, it cannot represent some specific necessary computations (e.g., inversion) and cannot represent iteration. The research on \lara's expressive ability in the relational algebra field, linear algebra field, and AI field is relatively complete. To make conclusions clear, we present propositions in Table~\ref{tab_conclusion}.

\begin{table}[ht]
\centering
\resizebox{0.65\textwidth}{!}{
\begin{tabular}{|cccc|}
\hline
Computation& $\lara(\Omega_=^{All})$& $\lara=\lara(\Omega_{All[n]}^{All})$& $\lara(\Omega_<^{\{+,\times\}})$\\\hline
$\sigma,\pi,\rho,\gamma$& $\checkmark$\cite{barcelo2022expressiveness}& $\checkmark$\cite{hutchison2016lara}& C\\\hline
$\cup,\cap,-,\div$& $\checkmark$\cite{barcelo2022expressiveness}& $\checkmark$\cite{hutchison2016lara}& C\\\hline
$\bowtie,\triangleright,\triangleleft$& C& C& C\\\hline
$+,\times,gemm,smm$& $\checkmark$\cite{barcelo2022expressiveness}& $\checkmark$\cite{hutchison2016lara}& C\\\hline
Pooling& $\checkmark$(Ours)& $\checkmark$(Ours)& ?\\\hline
Activation function& $\checkmark$(Ours)& $\checkmark$(Ours)& ?\\\hline
$Conv[n]$& $\times$\cite{barcelo2022expressiveness}& $\checkmark$\cite{barcelo2022expressiveness}& C\\\hline
$Conv$& $\times$\cite{barcelo2022expressiveness}& $\checkmark$\cite{barcelo2019expressiveness}& C\\\hline
$Inv[n]$& ?& $\checkmark$(Ours)& ?\\\hline
$Inv$& $\times$\cite{barcelo2022expressiveness}& $\times$(Ours)& ?\\\hline
$Det$& $\times$\cite{barcelo2022expressiveness}& $\times$(Ours)& ?\\\hline
$EinSum$& ?& $\checkmark$\cite{barcelo2019expressiveness}& C\\\hline
$RA_{Agg}$& $\checkmark$\cite{barcelo2022expressiveness}& C& C\\\hline
first-order logic& $\checkmark$\cite{barcelo2019arxiv}& C& C\\\hline
iteration& $\times$\cite{barcelo2022expressiveness}& $\times$\cite{barcelo2022expressiveness}& $\times$\cite{barcelo2022expressiveness}\\\hline
\end{tabular}}
\caption{This table shows the major conclusions of \lara's and its extensions' expressive ability. The references in the brackets are their first proofs. The letter ``C'' means that the proposition is a corollary. The ``?'' mark means the proposition is unsolved but irrelevant to this paper.}
\label{tab_conclusion}
\end{table}

\section{\iterlara: \lara~with iteration extension}
\lara~does not include any iteration and recursion operator. To extend \lara~with iteration, we define \iterlara. An iteration includes a condition expression and an iteration body. Before presenting the definition of \iterlara, we first analyze the syntax and semantics of \lara. Then we provide the entire definition of \iterlara~and some examples.

\subsection{\lara's syntax and semantics}
A \lara~expression is defined as an operator (union, join, or ext) with given input associative tables in the correct syntax, usually noted as $expr$ or $e$. The result of an expression must be a table. The formula $A\hourglass_+ B$, $A\bowtie_\times B$, and $\mathbf{ext}_{Sqrt} A$ are expressions. The formula $A\hourglass$ and $A\bowtie_- B$ are not expressions because $\hourglass$ requires two inputs, and the subtraction (-) function is not commutative.

The syntax of \lara~is shown as follows, where $e$, $e_1$, and $e_2$ stand for expressions, $\mbox{dom}(f)$ stands for the function's input domain of $f$.\\
$e::=A$, ($A$ is an associative table)\\
\color{white}.\color{black}$\hspace{1em}|\;\;e_1\hourglass_\oplus e_2$, ($K_{e_1}\cap K_{e_2}\neq\emptyset,\mbox{dom}(\oplus)=V_{e_1}=V_{e_2}$)\\
\color{white}.\color{black}$\hspace{1em}|\;\;e_1\hat{\bowtie}_\otimes e_2$, ($V_{e_1}\cap V_{e_2}\neq\emptyset,\mbox{dom}(\otimes)=V_{e_1}=V_{e_2}$)\\
\color{white}.\color{black}$\hspace{1em}|\;\;e_1\bowtie_\otimes e_2$, ($V_{e_1}\cap V_{e_2}\neq\emptyset,\mbox{dom}(\otimes)=V_{e_1}=V_{e_2}$)\\
\color{white}.\color{black}$\hspace{1em}|\;\;\mathbf{ext}_f e$, ($\mbox{dom}(f)=K_e\times V_e$)\\

The semantics of all \lara's expressions is shown as follows:\\
$A:\vec{k}\rightarrow\vec{v}$; for finite $\vec{k},A[\vec{k}]\neq=\omega$(default value); for all $A$, $A[\vec{0}]=\omega$; $A[\vec{}k]=\vec{v}$\\
$\oplus:(\vec{v}\times\vec{v})\rightarrow\vec{v}$\\
$\otimes:(\vec{v}\times\vec{v})\rightarrow\vec{v}$\\
$f:\vec{k}\times\vec{v}\rightarrow(\vec{k'}\rightarrow\vec{v'})$; for all $(\vec{k},\vec{v})$, for finite $\vec{k'}$, $f(\vec{k}:\vec{v})(\vec{k'})\neq\omega$\\
$e_1\hourglass_\oplus e_2:(K_{e_1}\times V_{e_1}\times K_{e_2}\times V_{e_2})\rightarrow(K_{e_1}\cap K_{e_2}\rightarrow V_{e_1}\cup V_{e_2})$\\
$e_1\hat{\bowtie}_\otimes e_2:(K_{e_1}\times V_{e_1}\times K_{e_2}\times V_{e_2})\rightarrow(K_{e_1}\cup K_{e_2}\rightarrow V_{e_1}\cap V_{e_2})$\\
$e_1\bowtie_\otimes e_2:(K_{e_1}\times V_{e_1}\times K_{e_2}\times V_{e_2})\rightarrow(K_{e_1}\cup K_{e_2}\rightarrow V_{e_1}\cup V_{e_2})$\\
$\mathbf{ext}_f e:(K_e\times V_e)\rightarrow((K_e\times\vec{k'})\rightarrow\vec{v'})$; $(\mathbf{ext}_f)e[K_e]=f(K_e:V_e)[\vec{k'}]$

\subsection{The iteration operator}
To define the iteration operator, we first define the expression function $F$ as $F:e_1\rightarrow e_2$, which means $F(e_1)=e_2$, namely $F$ is a replacement from $e_1$ to $e_2$. An expression function is used as an iteration body. For example, if $F(e)=e\bowtie C$, where $C$ is a constant table, then $F(F(e))=e\bowtie C\bowtie C$.

The semantics of function $F$ is as follows:\\
$F:e_1\rightarrow e_2$; $F:(\vec{k_1}\times\vec{v_1}\times\vec{k_2}\times\vec{v_2}\times\cdots)\rightarrow(K_{e_2}\times V_{e_2})$\\
We make the necessary explanation. The expression $e_1$ is an expression with a finite number of tables. Symbols $\vec{k_1},\vec{k_2},\cdots$ are their key attributes and symbols $\vec{v_1},\vec{v_2},\cdots$ are their value attributes. The expression $e_2$ is another expression with the given semantics $K_{e_2}\times V_{e_2}$, like a user-definable function. The function $F$ is a mapping from $e_1$ to $e_2$. Therefore, the semantics of $F$ is the above formula.

Now we define the iteration operator $\mathbf{iter}_F^{cond}$ as follows:
\begin{equation}\mathbf{iter}_F^{cond}e=\begin{cases}F(\mathbf{iter}_F^{cond}e),\mbox{ if Boole}[cond]\\e,\mbox{ otherwise}\end{cases}\nonumber\end{equation}

The condition expression $cond$ must be a Boolean expression. The feasibility of considering an associative table as a single numeric is explained in Section~\ref{sec53}. This operator is equivalent to a while-loop written in pseudo-code as follows:
\begin{equation}\mathbf{while}(cond)\{e=F(e)\};\;\;\mbox{result}=e\nonumber\end{equation}

The semantics of $cond$ and function $\mathbf{iter}_F^{cond} e$ is as follows, where $\mathbb{B}$ stands for Boolean space:\\
$cond:e_3\rightarrow\mathbb{B}$\\
$\mathbf{iter}_F^{cond} e:(\vec{k_1}\times\vec{v_1}\times\vec{k_2}\times\vec{v_2}\times\cdots)\times(K_{e_3}\times V_{e_3}\times\mathbb{B})\rightarrow(K_{e_2}\times V_{e_2})$\\

We name \lara~with this extended iteration operator as \iterlara. 

\subsubsection{\iterlara~with for-operator}
Especially when the number of iterations is a definite number $n$, we define another symbol as follows:
\begin{equation}\mathbf{for}_F^n e=\underbrace{F(F(\cdots F(e)\cdots))}_{n\mbox{ iterations of }F}\nonumber\end{equation}

This is equivalent to the following for-loop written in pseudo-code:
\begin{equation}\mathbf{for}(i=0;i\le n;i++)\{e=F(e)\};\;\;\mbox{result}=e\nonumber\end{equation}

In this restriction, we name \lara~with this extended iteration operator as $\iterlara^{Const}$.

If the number of iterations is not a constant but can be computed before entering the loop and will not change, we can also note it as follows:
\begin{equation}\mathbf{for}_F^{cond} e=\underbrace{F(F(\cdots F(e)\cdots))}_{cond\mbox{ iterations of }F}\nonumber\end{equation}
where the expression $cond$ is definite before entering the loop.

This is equivalent to the following for-loop written in pseudo-code:
\begin{equation}\mathbf{compute}\mbox{ }cond;\;\;\mathbf{for}(i=0;i\le cond;i++)\{e=F(e)\};\;\;\mbox{result}=e\nonumber\end{equation}

Specifically, when the expression $cond$ is the aggregation function $Count$, we name \lara~with this extended iteration operator as $\iterlara^{Count}$.
\subsection{Some explanations}\label{sec53}
The above definition mentioned that the condition expression $cond$ must be a Boolean value. However, an associative table $A$ with a single record of a single numeric value $a$ is still an associative table rather than a numeric one. How to apply a user-defined function $f$ on it (to let $f(A)=f(a)$)? This is feasible, as any user-definable function can only be used with an ext operator in an expression. For example, $\mathbf{ext}_{+f(A)}B$ can be considered as $\rho(\mathbf{ext}_f(A))\hourglass_+B$. We add rename operator $\rho$ to ensure that $K_{\rho(\mathbf{ext}_f(A))}=K_B$ and $V_{\rho(\mathbf{ext}_f(A))}=V_B$.

However, in the above example, the conversion from ext operator to union operator requires that the add (+) function is commutative and associative. How to apply this to another function that does not have this property?
This is also feasible by redefining a commutative and associative function and discarding part of its output using the ext operator when needed. For commutativity, if $h(x,y)\neq h(y,x)$, define $h'(x,y)=[[x,h(x,y)],[y,h(y,x)]]$. For associativity, if $h'(h'(x,y),z)\neq h'(x,h'(y,z))$, define $h''(x,y,z)=[[x,h'(h'(x,y),z)],[z,h'(x,h'(y,z))]]$ and $h''(x,y)=h''(x,y,\omega)$ where $\omega$ is the default value.

We use the slightly simplified symbols in the paper with the theoretical guarantee of the above explanation.

\subsection{Examples}
\textbf{Example}:\\
Given $F_1(e)=e\bowtie C$, where $C$ is a constant table. We have that
\begin{equation}\mathbf{for}_{F_1}^5(A)=A\bowtie C\bowtie C\bowtie C\bowtie C\bowtie C\nonumber\end{equation}\\
\textbf{Example}:\\
Given $F_2(a,e)=\mathbf{map}_{f_a} e\bowtie_\cup B$, where $\forall a,\forall x,f_a(x)=a\times x$. With given tables as vectors $A=[1,2]$, $B=[1]$, we compute the iterative expression $\mathbf{iter}_{F_2(Count(A),A)}^{A\hourglass_+ E_A<20} A$ as follows:\\
\textbf{Step 1:} $A=[1,2]$, $A\hourglass_+ E_A=3<20$, $A:=F_2(Count(A),A)\bowtie B=[1\times2,2\times2]\cup[1]=[2,4]\cup[1]=[2,4,1]$. \textbf{Step 2:} $A=[2,4,1]$, $A\hourglass_+ E_A=7<20$, $A:=F_2(Count(A),A)\bowtie B=[2\times3,4\times3,1\times3]\cup[1]=[6,12,3]\cup[1]=[6,12,3,1]$. \textbf{Step 3:} $A=[6,12,3,1]$, $A\hourglass_+ E_A=22>20$.

We get the result $\mathbf{iter}_{F_2(Count(A),A)}^{A\hourglass_+ E_A<20} A=[6,12,3,1]$. This procedure can be simulated by a while-loop in any computer language.

\section{\iterlara's expressive ability}
This section analyzes  \iterlara's expressive ability in three aspects. First, the minimum ability of \iterlara~is the same as that of \lara, supported by the equivalence of $\iterlara^{Const}$ and \lara. Second, $\iterlara^{Count}$ can represent matrix inversion and determinant. This conclusion shows that $\iterlara^{Count}$ is approximately enough as a unified computation abstraction model for big data, AI, scientific computing, and database fields. Third, the maximum ability of \iterlara~is the same as a Turing machine, which can represent any computation. We prove this proposition by representing all computations in another Turing complete algebra: BF language~\cite{mathis2011BF}.

\subsection{\iterlara~with constant times of iteration is equivalent to \lara}
\textbf{Proposition}:\\
$\lara=\iterlara^{Const}$.\\
\textbf{Proof}:\\
This proposition is trivial to prove. On the one hand, since $\iterlara^{Const}$'s only extended computation is $\mathbf{for}_F^n e$, where $n$ is a constant independent of the expression $e$ and function $F$, it can be replaced by the following definite expression: $\underbrace{F(F(\cdots F(e)\cdots))}_{n\mbox{ iterations of }F}$ in \lara. We have $\lara\subseteq\iterlara^{Const}$. On the other hand, $\iterlara^{Const}$ includes all operators in \lara, therefore its minimum expressive ability is not lower than \lara. We have $\lara\supseteq\iterlara^{Const}$.\hfill{$\square$}

\subsection{\iterlara~with aggregation can represent matrix inversion and determinant}\label{sec62}
\textbf{Proposition}:\\
$Inv\in\iterlara^{Count}$.\\
\textbf{Proof}:\\
Given a matrix represented as table $A:[[i,j]:v]$, whose size $n$ is unknown. We make $cond=A\hourglass_+E$, where $E:[i:v]$ is an empty table with only $i$ as its key attribute and the same value attribute as $A$. According to the definition, $cond=n$. Construct table $B:[[j,k]:w,B[j,1]=1$, similar to $B_1,\cdots,B_k$ in the proof of Proposition 3. We have $B=B_1$. Construct a user-definable function $f$ such that $\mathbf{ext}_f(B_i)=B_{i+1}$. Since $B_i$ has finite support, this is feasible. We make $F(A):=A\bowtie_\times B$. Therefore, we note $A'=\mathbf{for}_{F;B:=f(B)}^{cond}A=A\bowtie_\times B_1\bowtie_\times\cdots\bowtie_\times B_n$. The feasibility details of the expression ($\mathbf{for}_{F;B:=f(B)}^{cond}$) are explained in Section~\ref{sec63}.

Next step, we want to construct a similar function $d$ and table $C$ as in the proof of Proposition 3. For function $d$, it is easy for we can define an infinite quantity of functions $d_1(k_1:w),d_2([k_1,k_2]:w),\cdots$ and there must exist a proper function $d=d_n$. For table $C$, since we have table $A':[[k_1,\cdots,k_n]:w]$, we can get that $C=\mathbf{ext}_1(\sigma_{s}A')$ where $s(x)=\mathbf{true}$ when $x$ is a permutation. For the same reason, we have that $Det(A)=C\hourglass_+\mathbf{ext}_d(A')$ and $Inv(A)=A^*\bowtie_\times\Big(\mathbf{ext}_{reciprocal}Det(A)\Big)$.\hfill{$\square$}

\subsection{Some explanations}\label{sec63}
In the proof, we used this expression: $\mathbf{for}_{F;B:=f(B)}^{cond}A$. However, by definition of the iteration operator, we cannot change another table's value ($B$'s value) in the loop body. How to achieve the equivalent result? This is still feasible by the Cartesian product.

The Cartesian product is representable in \lara~and of course in \iterlara~by renaming ($\rho$) and relaxed join ($\bowtie$). According to the definition, any result of an expression is a table. The following expression:
\begin{equation}\mathbf{iter}_{F;e_1:=e_2}^{cond}e\nonumber\end{equation}
can be represented as follows:
\begin{equation}\pi_{e}(\mathbf{iter}_{F'}^{cond}(e\bowtie e_1)),\mbox{where }F'(e\bowtie e_1)=F(e)\bowtie e_2\nonumber\end{equation}
In the above formula, we leave out renaming operators. By defining a new expression function on the Cartesian product of tables, several statements can be unified as one statement. This expression is equivalent to the pseudo-code as follows:
\begin{equation}\mathbf{while}(cond)\Big\{e=F(e);\;\;e_1=e_2\Big\};\;\;\mbox{result}=e\nonumber\end{equation}

This method reports that a program written as several continuous statements without iteration is still equivalent to a single statement in \iterlara, a pure functional algebra. We use the slightly simplified symbols in the paper with the theoretical guarantee of the above explanation.

\subsection{\iterlara~is Turing complete}
We prove that \iterlara~without any restriction on user-definable functions, condition expressions, and iteration body is Turing complete. We demonstrate it by representing all operators in BF language.

\subsubsection{BF language}
BF language~\cite{mathis2011BF} is a minimum programming language created in 1993. The language consists of only eight operators. It is not designed for practical use but for research on language's extreme expressive ability. The prototype of BF language was proposed and proved Turing complete early in 1964~\cite{davis1966corrado}.

We present an equivalent version of this language. A program is written as a string. Each character of the string is one of its eight operators. The execution of a program requires a memory pointer $p_E$, an input pointer $p_I$, and an output pointer $p_O$. The code pointer goes character by character and executes the operator. The program halts when some specific condition is no longer satisfied. Its eight operators' are defined in Table~\ref{tab_BF}.

\begin{table}[ht]
\centering
\resizebox{0.65\textwidth}{!}{
\begin{tabular}{|lll|}
\hline
Operator& Description& \tabincell{l}{Equivalent\\pseudo-code}\\\hline
RShift $>$ & Pointer right shift by one& $++p_E$\\\hline
LShift $<$ & Pointer left shift by one& $--p_E$\\\hline
Inc + & Increase the pointed data by one& $++*p_E$\\\hline
Dec - & Decrease the pointed data by one& $--*p_E$\\\hline
Output \textbf{.} & Output the pointed data& $*(++p_O)=*p_E$\\\hline
Input \textbf{,} & Input the pointed data& $*(++p_E)=*p_I$\\\hline
LBrac [ & \tabincell{l}{Jump to the next RBrac operator\\if the pointed data is zero}& $\mathbf{while}(*p_E)\{$\\\hline
RBrac ] & \tabincell{l}{Jump to the last LBrac operator\\if the pointed data is non-zero}& $\}$\\\hline
\end{tabular}}
\caption{This table shows eight operators' definitions in BF language.}
\label{tab_BF}
\end{table}

\subsubsection{Proof of completeness}
\textbf{Proposition}:\\
\iterlara~is Turing complete.\\
\textbf{Proof}:\\
As is explained, several continuous statements without iteration are still equivalent to a single statement in \iterlara. As for the first six operators in BF language, we propose their equivalent expressions ($F_>$, $F_<$, $F_+$, $F_-$, $F_\textbf{.}$, and $F_\textbf{,}$). As for the last two operators that may include loops, we propose an equivalent expression of an entire loop body ($[F_1,\cdots,F_l]$, where $l\in\mathbb{N}$).

Before proposing equivalent expressions, we first define six associative tables. Table $E$, table $I$, and table $O$ stand for memory, input data, and output data, respectively. Table $P_E$, table $P_I$, and table $P_O$ stand for their pointer. Each pointer table has only one record at the start, and the record's value stands for the pointer's position. As Appendix~\ref{sec63} explains, six tables are equivalent to a single table by Cartesian product. Therefore, it is feasible to represent them in the pure functional \iterlara~algebra. See Table~\ref{tab_E}, \ref{tab_I}, \ref{tab_O}, \ref{tab_PE}, \ref{tab_PI}, and \ref{tab_PO}. The values in the brackets are default values.

\begin{minipage}[c]{0.3\textwidth}
\centering
\begin{tabular}{c|c}
 &[0]\\
entry$_E$& val$_E$\\\hline
0& 0\\
1& 1st memory bit\\
2& 2nd memory bit\\
$\vdots$& $\vdots$\\
\end{tabular}
\captionof{table}{$E$. The quantity of this table's records but 0 is the same as the operators' quantity in the initial memory data. The 0-th entry is a placeholder.}
\label{tab_E}
\end{minipage}
\begin{minipage}[c]{0.03\textwidth}
\centering
\begin{tabular}{c}
\end{tabular}
\end{minipage}
\begin{minipage}[c]{0.3\textwidth}
\centering
\begin{tabular}{c|c}
 &[0]\\
entry$_I$& val$_I$\\\hline
0& 0\\
1& 1st input bit\\
2& 2nd input bit\\
$\vdots$& $\vdots$\\
\end{tabular}
\captionof{table}{$I$. The quantity of this table's records but 0 is the same as the input data volume. The 0-th entry is a placeholder.}
\label{tab_I}
\end{minipage}
\begin{minipage}[c]{0.03\textwidth}
\centering
\begin{tabular}{c}
\end{tabular}
\end{minipage}
\begin{minipage}[c]{0.3\textwidth}
\centering
\begin{tabular}{c|c}
 &[0]\\
entry$_O$& val$_O$\\\hline
0& 0\\
\end{tabular}
\captionof{table}{$O$. This table is empty indeed. The 0-th entry is a placeholder.}
\label{tab_O}
\end{minipage}
\begin{minipage}[c]{\textwidth}
\centering
\begin{tabular}{c}
\end{tabular}
\end{minipage}
\begin{minipage}[c]{0.3\textwidth}
\centering
\begin{tabular}{cc|c}
 & &[$\omega$]\\
ptr$_E$& entry$_E$& dummy\\\hline
P& 0& $\omega$\\
\end{tabular}
\captionof{table}{$P_E$. The ``entry'' attribute can only be a natural number that stands for the pointer's position.}
\label{tab_PE}
\end{minipage}
\begin{minipage}[c]{0.03\textwidth}
\centering
\begin{tabular}{c}
\end{tabular}
\end{minipage}
\begin{minipage}[c]{0.3\textwidth}
\centering
\begin{tabular}{cc|c}
 & &[$\omega$]\\
ptr$_I$& entry$_I$& dummy\\\hline
P& 0& $\omega$\\
\end{tabular}
\captionof{table}{$P_I$. The ``entry'' attribute can only be a natural number that stands for the pointer's position.}
\label{tab_PI}
\end{minipage}
\begin{minipage}[c]{0.03\textwidth}
\centering
\begin{tabular}{c}
\end{tabular}
\end{minipage}
\begin{minipage}[c]{0.3\textwidth}
\centering
\begin{tabular}{cc|c}
 & &[$\omega$]\\
ptr$_O$& entry$_O$& dummy\\\hline
P& 0& $\omega$\\
\end{tabular}
\captionof{table}{$P_O$. The ``entry'' attribute can only be a natural number that stands for the pointer's position.}
\label{tab_PO}
\end{minipage}

We propose the operators' equivalent expression in \iterlara~as follows. Please note that we use some symbols from the extended relational algebra for simplification. As is explained in Appendix~\ref{sec63}, we also use several statements separated by ``;'' to represent one expression. Please note that $\pm1_k$ and $\pm1_v$ stand for increment and decrement on keys and values, respectively.
\begin{itemize}
\item $F_>:P_E\rightarrow\mathbf{ext}_{+1_k}P_E$,
\item $F_<:P_E\rightarrow\mathbf{ext}_{-1_k}P_E$,
\item $F_+:E\rightarrow\Big(\pi_{entry_E,val_E}\mathbf{ext}_{+1_v}(P_E\bowtie E)\Big)\hourglass(P_E\triangleleft E)$,
\item $F_-:E\rightarrow\Big(\pi_{entry_E,val_E}\mathbf{ext}_{-1_v}(P_E\bowtie E)\Big)\hourglass(P_E\triangleleft E)$,
\item $F_\textbf{.}:P_O\rightarrow\mathbf{ext}_{+1_k}P_O;\;\;O\rightarrow O\hourglass\Big(\pi_{entry_O,val_O}(\rho_{val_O/val_E}(P_E\bowtie E))\bowtie O\Big)$,
\item $F_\textbf{,}:P_I\rightarrow\mathbf{ext}_{+1_k}P_I;\;\;E\rightarrow E\hourglass\Big(\pi_{entry_E,val_E}(\rho_{val_E/val_I}(P_I\bowtie I))\bowtie E\Big)$,
\item Loop body $[F_1,\cdots,F_l]:T\rightarrow\mathbf{iter}_{F_1;\cdots;F_l}^{\pi_{val_E}(P_E\bowtie E)}(E\bowtie I\bowtie O\bowtie P_E\bowtie P_I \bowtie P_O)$.
\end{itemize}

Although the complete representation of the Cartesian product of tables is complicated, we proved the feasibility of representing all operators in BF language in \iterlara. Therefore, \iterlara~is Turing complete.\hfill{$\square$}

\section{OP: The operation count metric}
In this section, we define a theoretical metric of computation: Operation Count (OP, in short). We define OP metric on every operator in \lara. The OP of an algorithm can be computed with its \lara~representation by the combination of operators. The OP of an algorithm reflects the upper bound of its necessary computation amount. Specifically, the OP is its exact computation amount for some simple computations.

When an algorithm is represented in \lara, we get at least one computing method for it because all \lara~operators are computable. Any other computing method with more computation amount is inferior compared to it, but some optimizations may exist. Therefore, an algorithm's OP can be considered a reasonable upper bound of computation performance.

OP is also compatible with the existing computation metrics. For example, when the data are float point numbers in the high-performance computing field, OP is compatible with the FLOPs~\cite{institute1985ieee} metric in most conditions. In the algorithm analysis field that studies complexity, the OP's order of magnitude is consistent with time/space complexity metrics.

\subsection{The definition of OP}
We define the rule of counting on an expression's OP. If an expression is a given associative table, the OP is zero. If an expression combines several expressions with union, join, and ext operators, the OP of it equals the summation of the OP of several expressions and the OP of operators. The OP of an operator is decided by the operator's type and the volume of its table variables. We propose the OP's definition of union, strict join, and ext operators as follows:
\begin{enumerate}
\item $\mathbf{OP}(A\hourglass_\oplus B)=\mathbf{OP}(A)+\mathbf{OP}(B)+(n_{a,c}+n_{c,b})\times\mathbf{OP}(\oplus)$\\
\color{white}.\color{black}$\hspace{1em}\le\mathbf{OP}(A)+\mathbf{OP}(B)+(|a_1|\times\cdots\times|a_m|+|b_1|\times\cdots\times|b_p|)\times|c_1|\times\cdots\times|c_n|\times\mathbf{OP}(\oplus)$
\item $\mathbf{OP}(A\hat{\bowtie}_\otimes B)=\mathbf{OP}(A)+\mathbf{OP}(B)+n_{a,b,c}\times\mathbf{OP}(\otimes)$\\
\color{white}.\color{black}$\hspace{1em}\le\mathbf{OP}(A)+\mathbf{OP}(B)+|a_1|\times\cdots\times|a_m|\times|b_1|\times\cdots\times|b_p|\times|c_1|\times\cdots\times|c_n|\times\mathbf{OP}(\otimes)$
\item $\mathbf{OP}(\mathbf{ext}_f A)=\mathbf{OP}(A)+n_a\times\mathbf{OP}(f)$\\
\color{white}.\color{black}$\hspace{1em}\le\mathbf{OP}(A)+|a_1|\times\cdots\times|a_m|\times\mathbf{OP}(f)$
\end{enumerate}

In the above formulae, $A$ and $B$ are associative tables. Table $A$'s key attributes are $a_1,\cdots,a_m,c_1,\cdots,c_n$. Table $B$'s key attributes are $c_1,\cdots,c_n,b_1,\cdots,b_p$. The symbol $n_{attrs}$ represents the number of records with the key attributes $attrs$. The symbol $|attr|$ represents the number of records with the single key attribute $attr$. The OP of user-definable functions, such as $\mathbf{OP(\oplus)}$, $\mathbf{OP(\otimes)}$, and $\mathbf{OP(f)}$, should be predefined by the user before computing the OP. For example, if $\oplus=+$, we can define that $\mathbf{OP(\oplus)}=\mathbf{OP}(+)=1$.

As for the relaxed join, we have $A\bowtie_\otimes B=\mathbf{ext}_{f_A}\mathbf{ext}_{f_B}(A\hat{\bowtie}_\otimes B)$ according to its definition, where functions $f_A$ and $f_B$ are identity mapping functions that should have zero operation count. Therefore, the OP of the relaxed join equals that of the strict join, namely, $\mathbf{OP}(A\bowtie B)=\mathbf{OP}(A\hat{\bowtie} B)$.

\subsection{Examples}
\textbf{Example}:\\
We compute the OP of the classical matrix multiplication ($MatMul$) algorithm. The algorithm is represented in \lara~as $MatMul(A,B)=E\hourglass_+(A\bowtie_\times B)$ where $A:[[i,j]:v]$ is a matrix sized $M\times N$, $B:[[j,k]:v]$ is a matrix sized $N\times L$, and $E:[[i,k]:v]$ is an empty table. The operation count is computed as follows (note that $\mathbf{OP}(\mbox{constant table})=0$):\\
$\mathbf{OP}(MatMul(A,B))=\mathbf{OP}(E\hourglass_+(A\bowtie_\times B))$\\
\color{white}.\color{black}$\hspace{1em}=\mathbf{OP}(E)+\mathbf{OP}(A\bowtie_\times B)+(|E|+|A\bowtie_\times B|)\times OP(+)$\\
\color{white}.\color{black}$\hspace{1em}=0+\Big(\mathbf{OP}(A)+\mathbf{OP}(B)+n_{i,j,k}\times\mathbf{OP}(\times)\Big)+(0+|A\bowtie_\times B|)\times1$\\
\color{white}.\color{black}$\hspace{1em}=0+0+M\times N\times L\times1+M\times N\times L$\\
\color{white}.\color{black}$\hspace{1em}=2\times M\times N\times L$

This conclusion is consistent with the classical FLOPs $2MNL$ and the time/space complexity $O(MNL)$.

\subsection{OP in \iterlara~with aggregation}
We make trials to extend the OP metric to \iterlara. However, it cannot be well-defined in \iterlara~because Turing machine is proved to be undecidable. Instead, the OP extension in $\iterlara^{Count}$ is feasible, as each iteration number is decidable before entering the iteration body.

\begin{enumerate}
\item[\textbf{4.}] $\mathbf{OP}(\mathbf{for}_F^n e)=\mathbf{OP}(e)+\mathbf{OP}(F(e))+\mathbf{OP}(F(F(e)))+\cdots+\mathbf{OP}\Big(\underbrace{F(F(\cdots F(e)\cdots))}_{n\mbox{ iterations of }F}\Big)$
\item[\textbf{5.}] $\mathbf{OP}(\mathbf{for}_F^{cond} e)=\mathbf{OP}(e)+\mathbf{OP}(F(e))+\mathbf{OP}(F(F(e)))+\cdots+\mathbf{OP}\Big(\underbrace{F(F(\cdots F(e)\cdots))}_{cond\mbox{ iterations of }F}\Big)+\mathbf{OP}(cond)$
\end{enumerate}

The formulae 1-5 provide insight for developing automatic OP computing tools based on libraries built on \lara~and \iterlara, for example, LaraDB~\cite{hutchison2017laradb}. These tools can be considered theoretical references in the workload (algorithm) profiling research.

\section{Conclusion and future work}
This paper proposed a unified Turing complete algebraic model for big data, AI, scientific computing, and database fields. We first study the expressive ability of \lara: a key-value algebra in representing different operations on several existing studies. We conclude that the original \lara~language can represent all operations in the extended relational algebra and partial operations in the linear algebra, not including matrix inversion and determinant computation. To solve this problem, we propose an extension to \lara, named \iterlara. The study on \iterlara~concludes that its minimum ability is the same as that of \lara~and its maximum ability is the same as Turing machine. \iterlara~with some restrictions can still represent matrix inversion and determinant computation. We conclude that \iterlara~is a suitable algebraic model as the unified abstraction of general-purpose computation.

Some problems remain open for future work, such as the ``?'' marks in Table~\ref{tab_conclusion}. In other theoretical work, the expressive ability of \lara~plus some computations (such as $\lara+Inv$) is also essential. For practical research, it is worth implementing \iterlara~and the OP computing tools based on LaraDB~\cite{hutchison2017laradb} or other alternative libraries.

\bibliographystyle{ieeetr}
\bibliography{ref}

\end{document}